\documentclass[10pt,conference,a4]{IEEEtran}
\IEEEoverridecommandlockouts
\usepackage{cite}
\usepackage{amsmath,amssymb,amsfonts}
\usepackage{algorithmic}
\usepackage{graphicx}
\usepackage{textcomp}
\usepackage{xcolor}
\usepackage{braket}
\usepackage{qcircuit}
\usepackage{comment}
\usepackage{hyperref}
\usepackage{multirow}

\def\BibTeX{{\rm B\kern-.05em{\sc i\kern-.025em b}\kern-.08em
    T\kern-.1667em\lower.7ex\hbox{E}\kern-.125emX}}

\begin{document}
\bstctlcite{IEEEexample:BSTcontrol}
\title{Single Qubit State Estimation on NISQ Devices with Limited Resources and SIC-POVMs
}

\author{\IEEEauthorblockN{Cristian A. Galvis-Florez}
\IEEEauthorblockA{\textit{Department of Electrical}\\
\textit{Engineering and Automation,} \\
\textit{Aalto University}\\
Espoo, Finland \\
cristian.galvis@aalto.fi}
\and
\IEEEauthorblockN{Daniel Reitzner}
\IEEEauthorblockA{\textit{Quantum Algorithms and Software,} \\
\textit{VTT Technical Research Centre of Finland}\\
Espoo, Finland \\
daniel.reitzner@vtt.fi}
\and
\IEEEauthorblockN{Simo Särkkä}
\IEEEauthorblockA{\textit{Department of Electrical}\\
\textit{Engineering and Automation,} \\
\textit{Aalto University}\\
Espoo, Finland \\
simo.sarkka@aalto.fi}
}

\maketitle

\begin{abstract}
Current quantum computers have the potential to overcome classical computational methods, however, the capability of the algorithms that can be executed on noisy intermediate-scale quantum devices is limited due to hardware imperfections.
Estimating the state of a qubit is often needed in different quantum protocols, due to the lack of direct measurements.
In this paper, we consider the problem of estimating the quantum state of a qubit in a quantum processing unit without conducting direct measurements of it. 
We consider a parameterized measurement model to estimate the quantum state, represented as a quantum circuit, which is optimized using the quantum tomographic transfer function.
We implement and test the circuit using the quantum computer of the Technical Research Centre of Finland as well as an IBM quantum computer.
We demonstrate that the set of positive operator-valued measurements used for the estimation is symmetric and informationally complete.
Moreover, the resources needed for qubit estimation are reduced when direct measurements are allowed, keeping the symmetric property of the measurements.
\end{abstract}


\section{Introduction}
The development of quantum technology in recent years provides numerous advantages over its classical counterparts, making it promising for a wide range of applications \cite{Pelucchi2022, MacFarlane2003}. 
One of the applications of quantum technology concerns the development of quantum computers \cite{Nielsen2011}, whose computations are based on quantum bits (qubits). 
Current research in quantum computing includes the development of algorithms for problems that are computationally hard to solve on a classical computer and implement them on a quantum processing unit (QPU). 
These quantum devices have already shown advantages over their classical counterparts~\cite{Wu2021, Madsen2022}, however, limitations arising from their hardware imperfections must be considered in their analysis.

QPUs that have imperfections in quantum gates and measurement processes~\cite{Johnstun2021} are often called noisy intermediate-scale quantum (NISQ) devices~\cite{Preskill2018, Bharti2022}. 
A finite coherence time and imperfect performance of quantum gates limit the complexity of the algorithms that can be implemented on NISQ devices. 
Therefore, it is important to reduce these errors using quantum error correction algorithms \cite{Changjun2020, Jattana2020} or reduce the gates needed to complete an algorithm.
A specific set of basis gates can be implemented on a particular QPU. Usually, this corresponds to a set of single-qubit gates and a two-qubit gate, often a controlled-NOT (CNOT) gate. 
Two-qubit gates usually have error rates that are one or two orders of magnitude higher than single-qubit gates, making them a significant source of errors when executing quantum circuits \cite{Johnstun2021}.
To perfectly characterize and benchmark NISQ devices, we need to accurately estimate the state of a qubit.

\begin{figure}[t]
    \centering
    \includegraphics[width=0.99\linewidth]{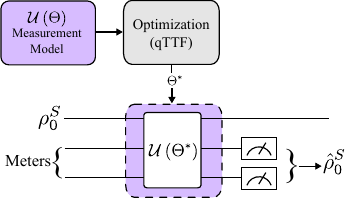}
    \caption{Concept of the paper. We aim to estimate the qubit $S$ in the state $\rho_0^S$ without measuring it directly. Instead, we use two ancillary qubits (meters) for that purpose. We propose a measurement model and optimize it using the quantum tomographic transfer function. Optimization provides the optimal interaction parameters $\Theta^*$. After the interaction, measurements of the meters allow us to compute the state estimate $\hat{\rho}_0^S$.}
    \label{fig:Concept_image}
\end{figure}
Estimating the quantum state of a quantum computer is an essential task to evaluate the performance of quantum algorithms \cite{Bendersky2009, James2001}.
A $d$-dimensional quantum state is identified by a density operator, characterized by $d^2 - 1$ real parameters, which are estimated using quantum state tomography methods~\cite{Dong2022, Chantasri2021} such as standard state tomography \cite{James2001}, using universal single observables \cite{Dariano2002, Mehmani2011, Wang2014}, machine learning techniques \cite{Lohani2021, Dominik2022}, or more recent classical shadow methods \cite{Huang2020, Stricker2022}. This process is described using a measurement model~\cite{Heinosaari2012}, characterized by a set of positive operator-valued measurement (POVM) elements, which are operators that describe the statistics of the measurement process. 
However, not every POVM can completely estimate a quantum state, in fact, the set of measurements that allow a complete estimation of quantum states are known as informationally complete POVM (IC-POVM)~\cite{Flammia2005}. 

The performance of the POVM elements that characterize an estimation process can be quantified and there are measurements that lead to more accurate quantum state estimates. 
In this work, the characterization of the measurements used to estimate a state is done using the quantum tomographic transfer function (qTTF)~\cite{Rehacek2015}.
A special class of IC-POVMs exhibits a symmetric attribute \cite{Renes2004}, which is often referred to as the symmetric informationally complete POVMs (SIC-POVMs).
The applications of SIC-POVM include quantum key distributions~\cite{Tavakoli2020} and the measurement of various quantum properties, including entanglement~\cite{Shang2018}.

In this paper, we start with the problem of estimating the state of one qubit in NISQ devices without conducting direct measurements of it (see Fig.~\ref{fig:Concept_image}). Direct implementations of tomography models such as \cite{Hacohen-Gourgy2016, Saavedra2019} in QPUs are usually impractical due to the depth of the circuit that simulates the interaction.
Making modifications to our previous work \cite{Galvis2023}, in this paper we propose a quantum circuit to estimate a qubit state measuring two ancillary qubits. 
A notable result of this paper is that the quantum circuit requires only two single qubit gates and two CNOT gates to estimate the qubit state, reducing gate errors.
Moreover, the POVM elements that characterize the estimation model are SIC-POVMs, meaning that the estimate's precision is independent of the initial state.
Finally, we show that when measurements over the estimated qubit are allowed, the circuit is reduced and requires only one CNOT gate, keeping the symmetric property of the POVM elements.

The paper is organized as follows. In Sec.~\ref{sec:Qubit_state_estimation_model} a parameterized measurement model is proposed. We describe the measurement model, how it is used to estimate the qubit, and the optimization of the estimator using the qTTF \cite{Rehacek2015}. 
In Sec.~\ref{sec:Quantum_Simulation_and_results} we propose a parametrized circuit that was optimized to perform qubit estimates and implemented on an IBM quantum computer and the Technical Research Center of Finland (VTT) quantum computer.
An analytical analysis of the measurement model in Sec.~\ref{sec:analytical_analysis} allowed us to determine that the optimization procedure led to a set of SIC-POVM elements. 
Also, we show how the quantum circuit can be simplified when measurements over the estimated qubit are allowed, keeping the symmetric property of the POVMs.
The work is concluded in Sec.~\ref{sec:Conclusions}.

\section{\label{sec:Qubit_state_estimation_model}Qubit State Estimation Model}
This section presents the estimation model used in this work. In Sec.~\ref{sec:Measurement_model} we construct the measurement model described by the interaction between the qubit and the meters. Then, in Sec.~\ref{sec:Qubit_state_estimator}, we describe the estimation method using the measurement results. Finally, Sec.~\ref{sec:characterization_of_estimators} shows the criteria for selecting the interaction parameters that allow better estimates.

\subsection{\label{sec:Measurement_model}Measurement model}
The tomography of a qubit $S$ is done by measuring the ancillary qubits $A$ and $B$, known as meters, providing information about the Bloch vector components of the qubit. Different interactions with the meters lead to partial~\cite{Peres1986, Perarnau2017} and complete~\cite{Saavedra2019} estimates of the Bloch vector.
We consider a one-qubit quantum state described by the density operator $\rho_0$.

The Hamiltonian $\mathcal{H}$, independent of time, describes the interaction between the system and the meters. 
The Hamiltonian defines the evolution operator $\mathcal{U}(\Theta) = e^{-iT\mathcal{H}/\hbar}$, that depends on the parameters $\Theta$ defined from the form of the Hamiltonian and the time parameter $T$.

After the interaction, we can prepare the estimation setup. For this purpose, the meters $A$ and $B$ are initialized in the state $\ket{0}$, while the system is in an arbitrary state $\rho_0^S$. The joint quantum state of the estimated qubit and the meters $S+A+B$ is $\rho_0 = \rho_0^S \otimes \ket{0^A}\bra{0^A}\otimes\ket{0^B}\bra{0^B}$. After the interaction we have 
\begin{equation}
    \rho_f(\Theta) = \mathcal{U}(\Theta)\left(\rho_0^S \otimes \ket{0^A}\bra{0^A}\otimes\ket{0^B}\bra{0^B}\right)\mathcal{U}^\dagger(\Theta).
\label{eq:state+S+A+B}
\end{equation}

Operator $\mathcal{U}(\Theta)$ can then be decomposed into the computational basis of subsystems A and B using the completeness relation $\sum_{i}\ket{i^A}\bra{i^A} = I$ as follows:
\begin{align*}
\mathcal{U}(\Theta)=&\sum_{i,k,j,l=1,0}\bra{k^Al^B}\mathcal{U}(\Theta)\ket{i^Aj^B}\otimes\ket{k^Al^B}\bra{i^Aj^B}\\
=&\sum_{i,k,j,l=1,0}\mathcal{U}_{kl,ij}(\Theta)\otimes\ket{k^Al^B}\bra{i^Aj^B}.
\end{align*}
Using the decomposition, state \eqref{eq:state+S+A+B} can be written as
\begin{equation}
\rho_f(\Theta) = \sum_{k,l = 0,1} M_{kl}(\Theta){\rho_0^S}M^{\dagger}_{kl}(\Theta)\otimes\ket{k^Al^B}\bra{k^Al^B},
\end{equation}
where we have used short-hand notation $M_{kl}(\Theta)=\mathcal{U}_{kl,00}(\Theta) = \braket{k^A l^B\mid\mathcal{U}(\Theta)\mid 0^A 0^B}$. Notice that the operators $M_{kl}$ with $k,l\in\{0,1\}$ act over the system $\rho_0^S$. Moreover, those are the Kraus operators representing the state change, given the outcomes $k$ and $l$ that are observed.

From the Kraus operators, we can now extract the POVM elements
\begin{equation}
    E_{kl} = M_{kl}^\dagger M_{kl}.
    \label{eq:POVM_elements_complete_estimation}
\end{equation}
The four possible outcomes of the measurement have probabilities
\begin{equation}
	p_{kl} = \operatorname{Tr}\left(M_{kl}\rho_0^S M_{kl}^\dagger\right) = \operatorname{Tr}\left(E_{kl}\rho_0^S\right),
 \label{eq:probabilities_kl}
\end{equation}
where $\rho_0^S$ is either a pure or mixed density operator of the initial state.
The final unnormalized state of the system after the measurement outcome $k,l$ is then 
\begin{equation*}
    \rho_{kl}^S = M_{kl}\rho_0^S M_{kl}^\dagger
\end{equation*}
In this paper, the measurement model is employed to estimate a qubit in a pure state $\rho_0^S = \ket{\psi_0^S}\bra{\psi_0^S}$.

\subsection{\label{sec:Qubit_state_estimator}Qubit state estimator}
Here we describe the methods used to estimate the initial state of the qubit $S$. The first one is the estimation by linear inversion (LI) \cite{Paris2004, Teo2015} and the second is the maximum likelihood estimator (MLE) \cite{Rehacek2007}. The latter always predicts positive semi-definite physical states but is computationally more expensive.
Recall the state $\rho_0^S$ of the qubit $S$. An equivalent description uses the column vector $s = \left( s_0 , s_1 , s_2 , s_3 \right)$. Here $s_0 = 1$ due to normalization, and the components $s_1, s_2$, and $s_3$ correspond to the $x,y,$ and $z$ components of the Bloch vector respectively.
The density matrix of the system can be written by 
\begin{equation}
    \rho^S_0 = \frac{1}{2}\sum_{\mu = 0}^{3} s_\mu \sigma_\mu,
\end{equation} 
where $\sigma_0$ is the two-dimensional identity matrix, while $\sigma_1$, $\sigma_2$, and $\sigma_3$ are the Pauli matrices $\sigma_x,\sigma_y$, and $\sigma_z$, respectively.

Recalling the probabilities of the four outputs in Equation~\eqref{eq:probabilities_kl} in terms of the POVM elements, and using the index $\nu \in \{0,1,2,3\}$ for the binary outcome pair $(k,l)$, we define matrix elements $T_{\nu\mu}$ as
\begin{equation}
	T_{\nu\mu} = \frac{1}{2} \operatorname{Tr} \left(E_\nu\sigma_\mu\right).
\end{equation}
Using \eqref{eq:probabilities_kl}, we can write the output probabilities as
\begin{equation}
    \mathbf{p} = \mathbf{T} \mathbf{s}.
\end{equation}

The measurement matrix is invertible if $\operatorname{det}(\mathbf{T})\neq 0$, a condition that relies upon the interaction parameters $\Theta$. if it is invertible, we can solve	$\textbf{s} = \mathbf{T}^{-1}(\Theta) \mathbf{p}.$
This allows us to estimate $\mathbf{s}$ using estimates of the probabilities $\hat{p}_{k} = n_{k}/N$. Then, the components of the Bloch vector can be estimated by
\begin{equation}
    \hat{s}_\mu= \sum_{\nu = 0}^4\left[\mathbf{T}^{-1}(\Theta)\right]_{\mu \nu} \hat{p}_\nu,
    \label{eq:spin_estimator}
\end{equation}
which is the LI estimator \cite{Paris2004, Teo2015}.

The linear estimates may lead to nonphysical states, which should be positive semi-definite states. This can be avoided by imposing the conditions~\cite{Fiurasek2001}
$\hat{s}_0 = 1$ and 
\begin{equation}
\left(1-\hat{r}_0\right) \hat{s}_\mu=\hat{r}_\mu
\end{equation}
for $\mu=1,2,3$, where
\begin{equation}
    \hat{r}_\mu=\sum_{\nu=1}^4 \frac{\hat{p}_\nu}{\check{p}_\nu(\Theta)} T_{\nu \mu}(\Theta).
\end{equation}
The parameters $\check{p}_k$ are not the estimators from the results of the experiment, they correspond to the estimate of the probabilities from the relation \eqref{eq:spin_estimator} using $\hat{\mathbf{s}}$ instead of $\mathbf{s}.$
The complexity of the measurement matrix makes the evaluation of these conditions difficult. However, the \textit{$R\rho R$ Algorithm} allows us to make the estimation numerically \cite{Rehacek2007}.

The estimation $\rho^{(n+1)}$ is computed from the previous estimate $\rho^{(n)}$ as
\begin{equation}        
{\rho}^{(n+1)} = \mathcal{N} \left[{R} \left({\rho}^{(n)}\right) {\rho}^{(n)} {R} \left({\rho}^{(n)}\right)\right],
	\label{eq:RpR_algorithm}
\end{equation}
where $\mathcal{N}[\cdot]$ denotes normalization to the unity trace of the corresponding operator, where
\begin{equation*}
    {R}=\sum_{\mu=0}^{3} \hat{r}_{\mu} \sigma_{\mu}.
\end{equation*}
In terms of the Bloch vector components, we have
\begin{equation}
    \hat{s}_\mu^{(n+1)}=\frac{2 \hat{r}_\mu^{(n)}-\hat{s}_\mu^{(n)} \hat{\gamma}^{(n)}}{2 \hat{r}_0^{(n)}+\hat{\gamma}^{(n)}},
\end{equation}
where
\begin{equation*}
    \hat{ \gamma}^{(n)} = \sum_{\mu = 1}^3 \left( \hat{r}_\mu^{(n)} \right)^2-\left( \hat{r}_0^{(n)} \right)^2.
\end{equation*}
This makes that $\hat{s}^{(n)}_0 = 1$ for every $n$. 
The initial state for the numerical implementation of this algorithm is the state 
$\hat{\mathbf{s}}^{(0)} =\left(1,0,0,0\right)$.

\subsection{\label{sec:characterization_of_estimators}Characterization of the estimator}
Every set of parameters $\Theta$ defines a different estimator. Its performance is evaluated using the Fisher information matrix \cite{Kay1993}.

The estimate is made using a quantum circuit that is executed $N$ times producing $n_m$ random samples of the measurement $m$. Then, we depose from a set of detections $\mathbf{n} = \{n_0, \dots,n_m\}$ where $\sum_m n_m = N$ that have a multinomial distribution 
\begin{equation}
	p(\textbf{n} \mid \textbf{p}) = N!\prod_{i=0}^{m}\frac{1}{n_i !}\left[p_{i}\right]^{n_i}.
	\label{eq:multinomial_prob_distribution}
\end{equation}
We can estimate the theoretical probabilities $p_i$ with $\hat{p}_i = n_i/N$ and $\lim_{N\rightarrow\infty}{\hat{p}_i} = p_i.$ Since the estimator is unbiased the Cramér-Rao relation $\operatorname{Cov}(\hat{\textbf{s}},\hat{\textbf{s}}')\geq \mathbf{J}^{-1}(\textbf{s})$ is satisfied \cite{Kay1993}. The matrix $\mathbf{J}$ is known as the Fisher information matrix, with components \cite{Teo2015}
\begin{equation}
	\mathbf{J}_{\mu\nu}(\textbf{s}) =- \mathbb{E}\left[\frac{\partial^2\ln\left[p(\textbf{n}\mid \mathbf{p})\right] }{\partial s_\mu\partial s_\nu}\right].
\end{equation}
Using $p(\textbf{n} \mid \textbf{p})$ described above \eqref{eq:multinomial_prob_distribution},
the Fisher information matrix components are proportional to the number of samples $N$.

We drop the $N$ term to consider only $\mathbf{F}$ which is related to the Fisher matrix by $\mathbf{J} = N \mathbf{F},$ with components
\begin{equation}
	\mathbf{F}_{\mu\nu} = \sum_{i = 0}^m \frac{1}{p_i}\frac{\partial p_i}{\partial s_\mu} \frac{\partial p_i}{\partial s_\nu}.
	\label{eq:Fisher_Matrix}
\end{equation}
The matrix $\mathbf{F}$ depends only on the state that will be estimated and the interaction parameters $\Theta$. For the measurement model described in Sec. \ref{sec:Measurement_model}, if we define the matrix $\mathbf{P} = \text{diag}(p_0,p_1,p_2,p_3),$ the components of the Fisher matrix take the form
\begin{equation}
    \mathbf{F}_{\mu\nu} = \left[ \mathbf{T}^{\top} \mathbf{P}^{-1} \mathbf{T} \right]_{\mu\nu}.
\end{equation}
The indexes $\mu\nu$ take the values associated with the estimated quantities, that is, $\mu,\nu \in \{1,2,3\}$ since $\partial p_i/\partial s_0 = 0$.

Now, we take the trace of the inverse of this matrix (which is related to the covariance matrix) to characterize the estimator. We name this the \textit{Fisher error parameter} and it has the form
\begin{equation}
	\Delta(\Theta, \textbf{s}) = \operatorname{Tr} \left(\mathbf{F}^{-1}(\Theta, \textbf{s})\right).
	\label{eq:Fisher_error_parameter}
\end{equation}
We are interested in making this error parameter small, but it depends on the initial state and the interaction parameters. To characterize the estimator independently of the initial state $\mathbf{s},$ we have to introduce the \textit{quantum tomographic transfer function} (qTTF) \cite{Rehacek2015}. This is defined as the average of the error parameter over all the possible initial states of the system. Without loss of generality, we will implement this tomography model to estimate pure quantum states of the form
\begin{equation}
\ket{\psi_0^S} = c_0 \ket{0} + c_1\ket{1}.
\label{eq:system_state}
\end{equation} 
Pure quantum states can be written in terms of the parameters $\xi = \{\alpha_1,\alpha_2\}$ as $c_0= e^{i\alpha_2}\cos\alpha_1$ and $c_1  = e^{- i\alpha_2}\sin\alpha_1$, where $\alpha_1\in [0,\pi/2],$ and $\alpha_2\in [0,\pi]$. Then, $\mathbf{s} = \mathbf{s}(\xi).$
With this parametrization, we are ready to define the qTTF as
\begin{equation}
	\operatorname{qTTF}(\Theta) = \frac{1}{V}\int_\xi \Delta(\Theta, \textbf{s}(\xi)) d\xi,
	\label{eq:qTTF}
\end{equation}
where $V = \int_\xi d\xi = \pi.$ As we can note, the qTTF does not depend on the initial state, only on the interaction parameters $\Theta$. Then, the best estimator will be the one whose interaction parameters satisfy
\begin{equation}
    \Theta^* = \arg \min _{\Theta} \operatorname{qTTF}( \Theta ).
\end{equation}

Once we have characterized the estimator, we can find the best set of parameters $\Theta^*$ to implement the measurement model.

\section{\label{sec:Quantum_Simulation_and_results}Quantum Simulation and results}
In this section, we construct a parametric circuit to estimate the qubit state and implement it on three different backends. In Sec.~\ref{sec:Parameterized_quantum_circuit} we describe the circuit and optimize it using the qTTF. Then, in Sec.~\ref{sec:Quantum_hardware_specifications} we specify the quantum hardware used for the implementations. Finally in Sec.~\ref{sec:results} we show and discuss the results.
\subsection{\label{sec:Parameterized_quantum_circuit}Parameterized quantum circuit}

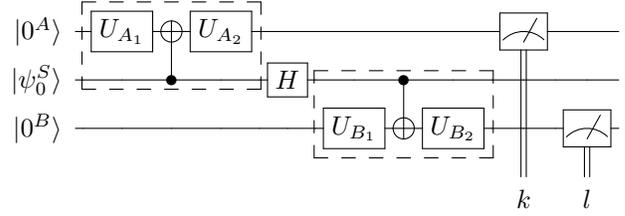
\begin{figure}[tb]
  \centering
    \begin{minipage}{8cm}
        \Qcircuit @C=0.3em @R=0.3em @!R { 
\push{\rule{2.2em}{0em}}& \lstick{\ket{0^A}}&\qw&\gate{{U}_{A_1}}&\targ &\gate{{U}_{A_2}} &\qw&\qw &\qw&\qw&\qw&\qw&\qw&\meter&\qw&\qw&\qw\\
&\lstick{\ket{\psi_0^S}}&\qw&\qw&\ctrl{-1}&\qw&\qw&\gate{H}&\qw&\qw&\ctrl{1}&\qw&\qw&\qw&\qw&\qw&\qw\\ &\lstick{\ket{0^B}}&\qw&\qw&\qw&\qw&\qw&\qw&\qw&\gate{{U}_{B_1}}&\targ&\gate{{U}_{B_2}}&\qw&\qw&\qw &\meter&\qw\\
&&&&&&&&&&&&&\dstick{k}\cwx[-3]& &\dstick{l}\cwx[-1]&&& \\
        &&&&&&&&&&&&&&&&&\relax
            \gategroup{1}{4}{2}{6}{.7em}{--} 
            \gategroup{2}{10}{3}{12}{.7em}{--}
}

    \end{minipage}
    \caption{Parametrized circuit to estimate the qubit $S$ performing measurements over the qubits $A$ and $B$. The U gates represent general rotations over the Bloch sphere. }
    \label{fig:optimal_circuit_structure}
\end{figure}

The interaction circuit proposed here consists of a parametrized circuit modified from our previous research \cite{Galvis2023}. This circuit shown in Fig. \ref{fig:optimal_circuit_structure} uses only two CNOT gates for performing a complete estimation of the qubit $S$.

The circuit depends on twelve parameters, three for each $U$ gate:
\begin{equation}
U(\vec\theta) = U(\theta, \phi, \lambda) =
\begin{pmatrix}
        \cos\frac{\theta}{2} & -e^{i\lambda}\sin\frac{\theta}{2} \\
        e^{i\phi}\sin\frac{\theta}{2} & e^{i(\phi+\lambda)}\cos\frac{\theta}{2}
    \end{pmatrix}.
    \label{eq:U_gate}
\end{equation}
Optimization of the qTTF over these parameters is accomplished using random initial parameters and the Nelder--Mead optimization method from SciPy \cite{2020SciPy-NMeth}. There are different sets of parameters that optimize the qTTF to the same value. A nice set of parameters $\Theta^*$ found by this optimization (with some detailed analysis in Sec. \ref{sec:analytical_analysis}) is
\begin{eqnarray*}
    \vec{\theta}^{*}_{A_1} = ( -\arccos\frac{1}{\sqrt{3}}, -\frac{\pi}{4}, 0)\quad
    \vec{\theta}^{*}_{A_2} = (0, 0, 0),\\
    \vec{\theta}^{*}_{B_1} = (0, 0, 0),\quad
    \vec{\theta}^{*}_{B_2} = (0, 0, 0),
\end{eqnarray*}

corresponding to an average error of $\bar{\Delta}\approx 8,$ which agrees with the minimal possible error of a POVM to estimate a single qubit pure state \cite{Rehacek2004}. 

\begin{figure}[tb]
    \centering
    \includegraphics[width = 0.95\linewidth]{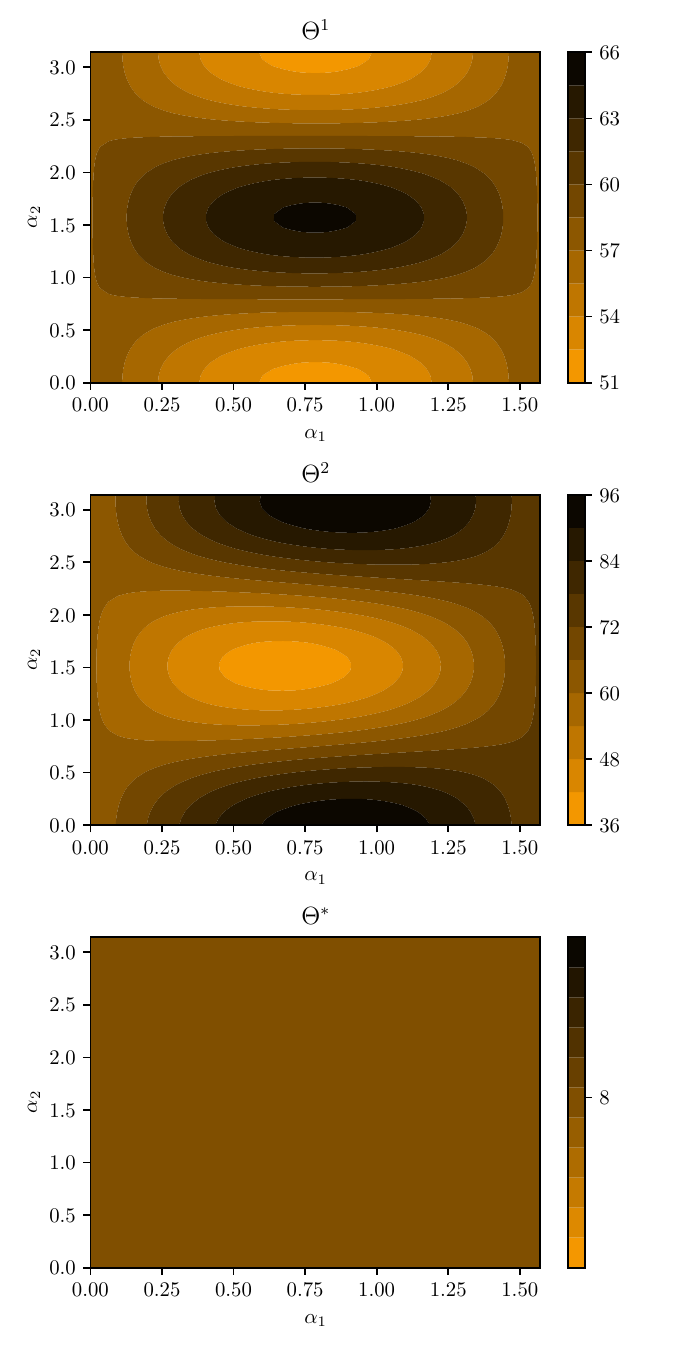}
    \caption{Error parameter in Equation~\eqref{eq:Fisher_error_parameter} as a function of the state parameter $\xi = \{\alpha_1, \alpha_2\}$ for different sets of parameters $\Theta.$ The values $\Theta^1 = \{\vec{\theta}_{A_1}=(2.01, 1.32, 0.51),$ $\vec{\theta}_{A_2}=(0.00, 4.60, 4.24),$ $\vec{\theta}_{B_1}=(0.95, 5.56, 4.48),$ $\vec{\theta}_{B_2}=(2.35, 0.54, 0.60)\}$ and $\Theta^2 = \{\vec{\theta}_{A_1}=(2.26, 2.48, 5.25),$ $\vec{\theta}_{A_2} = (3.07, 5.18, 6.07),$ $\vec{\theta}_{B_1} = (1.47, 5.88, 5.3),$ $\vec{\theta}_{B_2} = (1.18, 1.09,3.68)\}$ where randomly chosen. The final graph corresponds to the already reported optimal values $\Theta^*.$ }
    \label{fig:Error_every_state}
\end{figure}

To study the estimation error for different initial states, in Fig.~\ref{fig:Error_every_state}, we graphed the Fisher error \eqref{eq:Fisher_error_parameter} as a function of $\alpha_1$ and $\alpha_2$.
The parameters $\Theta^1$ and $\Theta^2$ where randomly chosen while $\Theta^*$ represents the optimal parameters. 
The magnitude of the error value for the parameters $\Theta^{1,2}$ is larger with respect to the graph of the optimal parameters.
Also, between graphs $\Theta^1$ and $\Theta^2$, there are parameters that reduce the estimation error for specific initial states showing that some sets of POVMs are not symmetrical.  
Finally, the most notable result, is that there are no variations of the error for any initial state when the optimal parameters $\Theta^*$ are chosen. 
This implies that the accuracy of the estimates is independent of the initial state, a characteristic of SIC-POVMs. 

\subsection{\label{sec:Quantum_hardware_specifications}Quantum hardware specifications}
\begin{figure}[tb]
    \centering
    \includegraphics[width = 0.5\linewidth]{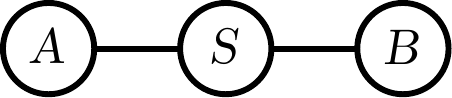}
    \caption{Choice of qubit order for the implementation of the circuit on a QPU.}
    \label{fig:qubit_topology}
\end{figure}

The implementation of the circuit in Fig.~\ref{fig:optimal_circuit_structure} was made using Qiskit \cite{Qiskit}. The topology of the QPU is important for the execution of the circuit. We selected three qubits that are beside each other, placing the qubit $S$ in the middle of the meter qubits as in Fig.~\ref{fig:qubit_topology}.
Changing the order of the qubits could imply the need for additional CNOT gates to execute the circuit. 

The implementation of our method with the IBM quantum computer \cite{Devitt2016, ibm-quantum} was made using the \texttt{ibmq\_lima} QPU.
This is a five-qubit quantum processor of the type Falcon r4T. 
For the implementation, we used the qubits $0,$ $1,$ and $2$ for $A$, $S,$ and $B$ respectively. 
The basis gates of this processor are $\operatorname{CNOT}$ (controlled-NOT), $\operatorname{ID}$ (identity), $\operatorname{RZ}$ (Rotation around $z$ axis), $\operatorname{X}$ (NOT), and $\operatorname{SX}$ ($\sqrt{\operatorname{X}}$).
Table~\ref{tab:QPUs_details} shows the relaxation time ($T_1$), the dephasing time ($T_2$), the single-qubit gates error ($\epsilon_1$), and the two-qubit gates error ($\epsilon_{CX}$).

Our method was also implemented on the VTT quantum computer Helmi \cite{FiQCI}. This is a five-qubit quantum processor with basis gates $\operatorname{CZ}$ (controlled-$\operatorname{Z}$), $ID$ (identity), $\operatorname{RX},$ $\operatorname{RY},$ and $\operatorname{RZ}$ (rotations around the $x,y,z$ axis respectively). We used the qubits $\operatorname{QB1},$ $\operatorname{QB3},$ and $\operatorname{QB2}$ for the qubits $A$, $S$, and $B$ respectively. Table~\ref{tab:QPUs_details} show the relaxation time ($T_1$), the dephasing time ($T_2$), the single-qubit gates error ($\epsilon_1$), and the two-qubit gates error ($\epsilon_{CZ}$).
\begin{table}[tb]
\caption{\label{tab:QPUs_details} Calibration details of the \texttt{ibmq\_lima} and Helmi quantum processing units by April 2023.}
\begin{center}
\begin{tabular}{lcccp{28mm}}
\hline \hline
\multicolumn{5}{c}{IBMQ}\\\hline
Qubit & $T_1(\mu \mathrm{s})$ & $T_2(\mu \mathrm{s})$ & $\epsilon_1$ & $\epsilon_{CX}$ \\\hline
0 & 40.09 & 105.23 & $5.80 \times 10^{-4}$ & $0\leftrightarrow 1 :7.96 \times 10^{-3}$ \\
1 & 75.77 & 133.43 & $3.04 \times 10^{-4}$ & $1\leftrightarrow 2 :6.51 \times 10^{-3}$\\
2 & 27.92 &  89.73 & $5.94 \times 10^{-4}$ & \\
\hline\hline
\multicolumn{5}{c}{Helmi}\\\hline
 Qubit & $T_1(\mu \mathrm{s})$ & $T_2(\mu \mathrm{s})$ & $\epsilon_1$ & $\epsilon_{CZ}$ \\\hline
$\operatorname{QB1}$ & 35.07 & 20.93 & $1.96 \times 10^{-3}$ & $\operatorname{QB1}\leftrightarrow \operatorname{QB3} : 0.0814 $ \\
$\operatorname{QB2}$ & 29.84 & 20.00 & $5.25 \times 10^{-3}$ & $\operatorname{QB2}\leftrightarrow \operatorname{QB3} : 0.0664 $ \\
$\operatorname{QB3}$ & 49.82 &  14.15 & $1.61 \times 10^{-3}$ &\\ 
\hline \hline
\end{tabular}
\end{center}
\end{table}

\subsection{\label{sec:results}Results and Discussion}
Demonstrations of our estimation method were performed by initializing the qubit $S$ over the six eigenstates of the Pauli matrices 
\begin{equation}
\begin{split}
X&= \left\{ \ket{\psi_{z_0}}, \ket{\psi_{z_1}}, \ket{\psi_{x_0}}, \ket{\psi_{x_1}}, \ket{\psi_{y_0}}, \ket{\psi_{y_1}}\right\}\\
&= \left\{|0\rangle, |1\rangle, \frac{|0\rangle+|1\rangle}{\sqrt{2}}, \frac{|0\rangle-|1\rangle}{\sqrt{2}}, \frac{|0\rangle+i|1\rangle}{\sqrt{2}}, \frac{|0\rangle-i|1\rangle}{\sqrt{2}}\right\},
\end{split}
\end{equation}
estimating each of these states 5 times, with 1024 shots per circuit. The demonstrations were executed on three different backends, the qiskit classical simulator \texttt{qasm\_simulator} (QASM), the IBM quantum computer \texttt{ibmq\_lima} (IBMQ), and the VTT quantum computer (Helmi). We consider the LI estimator and the $R\rho R$ algorithm, then, we graphed the purity $\gamma = \text{Tr}(\hat{\rho}^2)$ of the estimates for every state to find non-physical states.

\begin{figure}[tb]
    \centering
    \includegraphics[width = 0.99\linewidth]{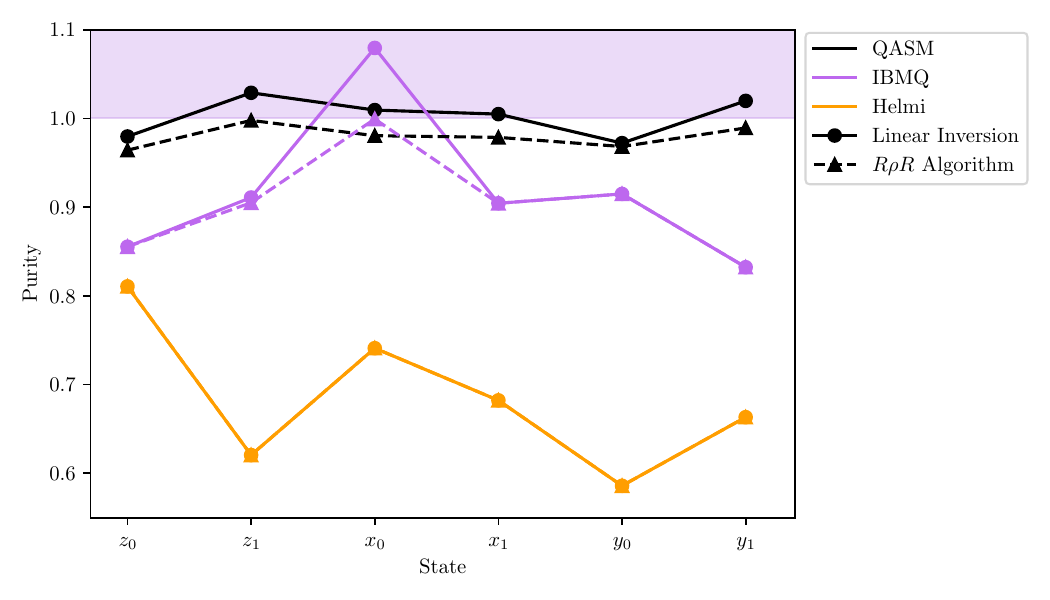}
    \caption{Purity of the averaged estimates on QASM (black), IBMQ (purple), and the  Helmi (orange) backends. The linear inversion estimates are shown with a solid line while the $R\rho R$ algorithm estimates are shown with a dashed line. The shaded region envelopes the non-physical estimates.}
    \label{fig:purity}
\end{figure}

Fig.~\ref{fig:purity} shows that there are states where the LI algorithm estimates states such that $\gamma > 1$, implying that is a non-physical state. The graph shows that the non-physical predictions took place when the state was near to being pure. For low purity estimates, the LI and the $R\rho R$ algorithm results are the same. For the following computations, we employ the $R\rho R $ algorithm to avoid this problem. 

The results of the demonstrations for state estimation are shown in Table \ref{tab:qpu_total_s_results_opt}. 
\begin{table}[tb]
\caption{\label{tab:qpu_total_s_results_opt} Average estimates $\hat{\vec{s}} = (\hat{s}_x, \hat{s}_y, \hat{s}_z)$ of the Bloch vector $\vec{s} = (s_x,s_y,s_z)$ using the QASM, IBMQ, and Helmi backends. The demonstrations were implemented using the circuit described in Fig.\ref{fig:optimal_circuit_structure}. The circuit was executed $5$ times with $1024$ shots per execution.}
\begin{center}
\begin{tabular}{rrrr}
\hline\hline
&\multicolumn{3}{c}{QASM}\\\cline{2-4}
State   &\multicolumn{1}{c}{ $\hat{s}_x$}&\multicolumn{1}{c}{ $\hat{s}_y$}     & \multicolumn{1}{c}{ $\hat{s}_z$}   \\
    \hline
$\ket{\psi_{z0}}$ &  $ -0.05\pm 0.03$ & $ 0.06\pm 0.04$ & $ 0.96\pm 0.04$ \\
$\ket{\psi_{z1}}$ &  $ 0.0\pm 0.06$ & $ 0.04\pm 0.03$ & $ -0.997\pm 0.002$ \\
$\ket{\psi_{x0}}$ &  $ 0.98\pm 0.02$ & $ 0.04\pm 0.07$ & $ 0.01\pm 0.05$ \\
$\ket{\psi_{x1}}$ &  $ -0.98\pm 0.02$ & $ -0.0\pm 0.04$ & $ 0.02\pm 0.03$ \\
$\ket{\psi_{y0}}$ &  $ 0.01\pm 0.02$ & $ 0.97\pm 0.05$ & $ -0.02\pm 0.06$ \\
$\ket{\psi_{y1}}$ &  $ 0.02\pm 0.04$ & $ -0.99\pm 0.02$ & $ 0.06\pm 0.05$ \\
    \hline
&\multicolumn{3}{c}{IBMQ}\\\cline{2-4}
State &\multicolumn{1}{c}{ $\hat{s}_x$}&\multicolumn{1}{c}{ $\hat{s}_y$}     & \multicolumn{1}{c}{ $\hat{s}_z$}   \\
\hline
$\ket{\psi_{z0}}$ &  $ 0.09\pm 0.02$ & $ -0.04\pm 0.02$ & $ 0.84\pm 0.06$ \\
$\ket{\psi_{z1}}$ &  $ 0.03\pm 0.04$ & $ 0.01\pm 0.04$ & $ -0.9\pm 0.07$ \\
$\ket{\psi_{x0}}$ &  $ 0.993\pm 0.001$ & $ -0.04\pm 0.04$ & $ -0.1\pm 0.02$ \\
$\ket{\psi_{x1}}$ &  $ -0.9\pm 0.04$ & $ -0.04\pm 0.08$ & $ -0.05\pm 0.1$ \\
$\ket{\psi_{y0}}$ &  $ 0.2\pm 0.04$ & $ 0.89\pm 0.04$ & $ -0.03\pm 0.03$ \\
$\ket{\psi_{y1}}$ &  $ 0.04\pm 0.05$ & $ -0.81\pm 0.05$ & $ -0.09\pm 0.06$ \\
\hline
&\multicolumn{3}{c}{Helmi}\\\cline{2-4}
State   &\multicolumn{1}{c}{ $\hat{s}_x$}&\multicolumn{1}{c}{ $\hat{s}_y$}     & \multicolumn{1}{c}{ $\hat{s}_z$}   \\
\hline
$\ket{\psi_{z0}}$ &  $ 0.08\pm 0.04$ & $ -0.12\pm 0.04$ & $ 0.77\pm 0.06$ \\
$\ket{\psi_{z1}}$ &  $ -0.03\pm 0.02$ & $ -0.02\pm 0.07$ & $ -0.49\pm 0.07$ \\
$\ket{\psi_{x0}}$ &  $ 0.68\pm 0.05$ & $ 0.11\pm 0.03$ & $ 0.05\pm 0.05$ \\
$\ket{\psi_{x1}}$ &  $ -0.47\pm 0.04$ & $ -0.26\pm 0.06$ & $ 0.28\pm 0.03$ \\
$\ket{\psi_{y0}}$ &  $ 0.16\pm 0.05$ & $ 0.35\pm 0.06$ & $ 0.14\pm 0.06$ \\
$\ket{\psi_{y1}}$ &  $ 0.04\pm 0.09$ & $ -0.51\pm 0.08$ & $ 0.26\pm 0.07$ \\
\hline\hline
\end{tabular}
\end{center}
\end{table}
The QASM simulator results show what we would expect from a noiseless quantum computer with a limited number of shots. The estimates computed from the QASM simulator are the most accurate and precise. The demonstrations on a real quantum computer are less accurate than the QASM simulator as would be expected, due to imperfections on the quantum computer gates as well as the state preparation and measurement (SPAM) \cite{Jattana2020, Johnstun2021}. 
The estimates can be compared by computing the fidelity between the initial state of the system $\rho_0^S$ and the estimated state $\hat{\rho}_0^S$:
\begin{equation}
  \mathcal{F} (\rho_0^S,\hat{\rho}_0^S) = \left[\operatorname{Tr}\left(\sqrt{\sqrt{\rho_0^S} \hat{\rho}_0^S \sqrt{\rho_0^S}}\right)\right]^2.
\end{equation}

\begin{figure}[tb]
    \centering
    \includegraphics[width = 0.99\linewidth]{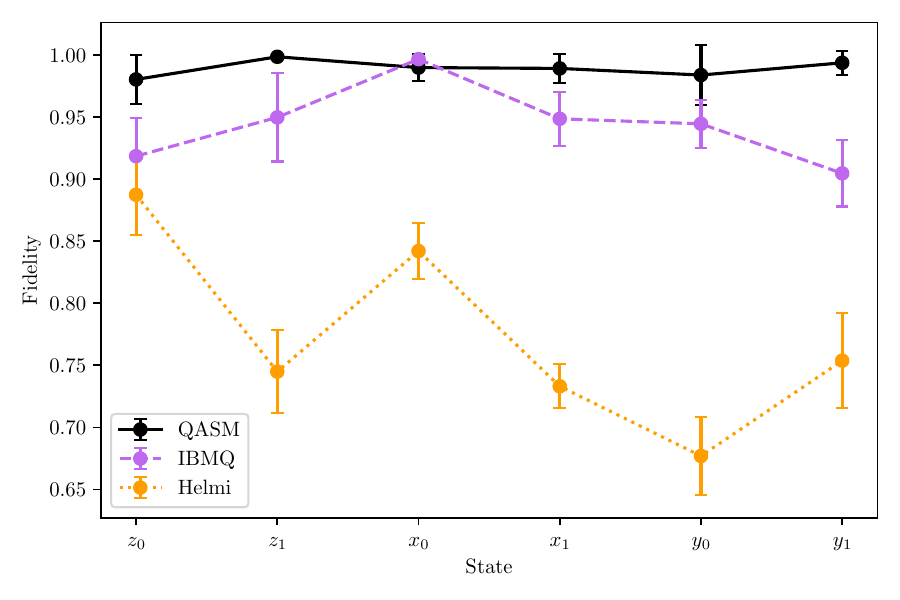}
    \caption{Fidelity of the averaged estimates on a QASM simulator (black solid), the IBMQ (purple dashed), and Helmi quantum computers (orange dotted).}
    \label{fig:fidelity}
\end{figure}
Fig. \ref{fig:fidelity} shows the average fidelity of the estimates for each state. The QASM simulator shows accurate estimates with fidelities near the unit while estimates of the real quantum computers have lower fidelities, as expected from noisy hardware. Estimates computed on IBMQ show fidelities between $0.90$ and $0.99$, greater than the estimates computed on Helmi which are between $0.68$ and $0.89$, showing more accuracy on the IBMQ quantum computer.
This can be expected from the coherence times and gate error rates of the QPUs shown in Table~\ref{tab:QPUs_details}. 
This effect can also be seen as the fidelity of the estimates executed on Helmi QPU tend to vary more between different states.

\section{\label{sec:analytical_analysis}Analytical analysis}
After the implementation, we decided to perform a theoretical analysis of the quantum circuit. The previous results suggested that the measurements could be symmetric, which is demonstrated in Sec.\ref{sec:Undestanding_the_circuit}. Also, we noticed that the circuit can be simplified when measurements of the $S$ qubit are allowed, shown in Sec.\ref{sec:Further_simplification}.

\subsection{\label{sec:Undestanding_the_circuit}Understanding the circuit}
The optimization of the qTTF is Haar-uniform over all the states. This suggests that also the POVM elements that are achieved will have symmetry similar to that of the state space. Moreover, optimization should provide an extremal POVM. These observations give us the possibility to analytically analyze the optimized circuit.

The first point to notice is the fact that the operators $E_{kl}$ are SIC-POVMs \cite{Renes2004}. This means, with $E_{kl}=\frac{1}{d}\Pi_{kl}$ (trace is $1/d$), satisfy
\begin{equation}
    \mathrm{Tr}(\Pi_{kl}\Pi_{mn})=\frac{d\delta_{km}\delta_{ln}+1}{d+1},
    \label{eq:SIC-POVM_condition}
\end{equation}
where $d=2$ is the dimension of the system and $\Pi_{kl}$ are 1-dimensional projectors corresponding to outcome indices $(k,l)$. Indeed, to numerical precision, the traces of the POVM elements are $1/2$ and their computed overlaps are $1/12$. This is a welcomed result as SIC-POVMs are considered to be optimal for state estimation \cite{Haah2017, Garcia-Perez2021}.

To better understand the formation of the SIC-POVM, we can formally split the circuit into three parts, measurement using subsystem $A$, intermediate Hadamard transform, and measurement using subsystem $B$. This distinction is also denoted in Fig.~\ref{fig:optimal_circuit_structure}. Under considered splitting, we can obtain POVM elements for these parts in a similar way as was done above for the whole circuit which leads to Equation~\eqref{eq:POVM_elements_complete_estimation}.
 
The measurement on subsystem $B$ turns out to be phased projections to either state $\ket{0}$ or $\ket{1}$. Firstly, as the measurement is final and we do not care about the state change, we can replace it with a simple $z$-measurement. This simplifies the measurement considerably, as there is no need for the rotations $U_{B_1}$ and $U_{B_2}$. Understandable, because if the last measurement would not be projective, its sharpening would lead to a better resolution, which should not be the case for optimal measurement. Another point to notice is that once we remove the unitaries on the $B$ system, we are only left with the CNOT coupling gate. If we accept the possibility of measuring system $S$ directly (which might not always be the case), we can furthermore identify subsystem $B$ with the system $S$ and perform the $z$-measurement directly on $S$.

 \subsection{\label{sec:Further_simplification}Further simplification}

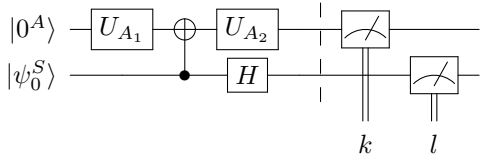
\begin{figure}[tb]
\centering
    \mbox{
        \Qcircuit @C=0.8em @R=0.2em @!R{
   \push{\rule{2em}{0em}}& \lstick{\ket{0^A}}&\gate{{U}_{A_1}}&\targ    &\gate{{U}_{A_2}}&\qw\barrier[0em]{1}&\qw&\meter                 & \qw                     & \qw \\
   & \lstick{\ket{\psi_0^S}}&\qw                            &\ctrl{-1}&\gate{H}                       &\qw                &\qw&\qw                    & \meter                   & \qw \\
   &&                                           &      &                           &              &&\dstick{k} \cwx[-2] & \dstick{l}\  \cwx[-1] & \\
        &&&&&&&&}
    }
    \caption{Simplified circuit for estimation of qubit $S$ when direct measurements are allowed.}
    \label{fig:simplified_qubit_circuit}
\end{figure}

Now let us consider a simplified circuit on subsystems $S$ and $A$ that is composed of two generic unitaries $U_{A_1}$ and $U_{A_2}$ on $A$, preceding and following the CNOT between $S$ and $A$, which is followed by a Hadamard on $S$ and measurement on $S$ (see Fig.~\ref{fig:simplified_qubit_circuit}). Parametrizing the two unitaries by angles $(\theta_1,\phi_1,\lambda_1)$ and $(\theta_2,\phi_2,\lambda_2)$ respectively, using the unitary parametrization in Equation~\eqref{eq:U_gate}.

The POVM elements of the circuit are
\begin{equation}
    E_{kl}=\frac{1}{4}\left[
        (1-kb)I+(lx_1-klx_2)\sigma_x+kly\sigma_y+kz\sigma_z
    \right],
\end{equation}
where
\begin{align}
    b &= \sin\theta_1\cos\phi_1\sin\theta_2\cos\lambda_2, \\
    x_1 &= \sin\theta_1\cos\phi_1, \\
    x_2 &= \sin\theta_2\cos\lambda_2, \\
    y &= \cos\theta_1\sin\theta_2\sin\lambda_2 - \sin\theta_1\sin\phi_1\cos\theta_2, \\
    z &= \cos\theta_1\cos\theta_2 + \sin\theta_1\sin\phi_1\sin\theta_2\sin\lambda_2.
\end{align}
We can notice the independence on $\lambda_1$ and $\phi_2$, which can be set arbitrarily. In order for the elements to form SIC-POVM we want to fulfill both Equation~(\ref{eq:SIC-POVM_condition}) and the normalization
\begin{equation}
    \mathrm{Tr}(E_{kl})=\frac{1-kb}{2} = \frac{1}{2}.
\end{equation}
The latter condition leads to the requirement
\begin{equation}
    b = \sin\theta_1\cos\phi_1\sin\theta_2\cos\lambda_2 = 0,
\end{equation}
which leads to four (not necessarily disjoint) conditions for the four relevant angles. Simple evaluation of cases $\sin\theta_1=0$ and $\cos\phi_1=0$ does not give a solution, because the Bloch vectors representing the directions of the measurements of the four POVM elements are co-planar and the normalization (\ref{eq:SIC-POVM_condition}) cannot be fulfilled. The case of $\cos\lambda_2=0$ is a non-trivial case for which the numerical optimization given above is an example of. While this case may provide a whole class of solutions, a more interesting case is the last one, where $\sin\theta_2=0$. We will not perform a full analysis, but rather a simplified analysis that leads to a solution that is of interest.

The normalization condition does not depend on $\lambda_2$, which we can set arbitrarily. We can thus go for the choice $(\theta_2,\phi_2,\lambda_2)=(0,0,0)$ which gives $U_{A_2}=I$.  This choice reduces the number of non-trivial gates, which is important from a practical point of view.

Furthermore, from the normalization condition we get that $\cos2\theta_1=\pm\frac{1}{3}$ and
\begin{equation}
    \mid\cos^2\theta_1\pm\sin^2\theta_1\cos 2\phi_1\mid=\frac{1}{3}.
\end{equation}
The two conditions together require $\cos^2\theta_1=\frac{1}{3}$ and $\cos2\phi_1=0$. To obtain the same POVM elements as in the numerical optimization above, we can choose
\begin{equation}
    \theta_1 = -\arccos\frac{1}{\sqrt{3}},\qquad\phi_1=-\frac{\pi}{4},\qquad\lambda_1=0.
    \label{eq:optimal_values}
\end{equation}
This leads to the final form for the POVM elements,
\begin{equation}
    E_{kl}=\frac{1}{4}\left[
        I-\frac{1}{\sqrt{3}}(l\sigma_x+kl\sigma_y-k\sigma_z)
    \right].
\end{equation}
Which corresponds to a set of SIC-POVMs. In fact, optimization of the qTTF in Equation \eqref{eq:qTTF} where the initial values are near to the values in Equation \eqref{eq:optimal_values}, the same solution is found.
This implies that a unique ancilla qubit and a circuit with one CNOT gate are enough for performing estimates of a qubit with SIC-POVMs. After peer review, we found a recent work about optimal measurement operators in fermion-to-qubit mapping \cite{Jiang2020}. That work suggests a similar circuit for optimal state estimation to the one reported in this paper, however, the interaction is deduced using a ternary trees technique. In Ref. \cite{Garcia-Perez2021} the authors explore adaptative SIC tomography using a variational circuit and the scope of SIC-POVMs against other estimation techniques. 

The estimation model reported here demonstrates being efficient for single-qubit state estimation protocols in quantum processors. The unitary operation of the quantum circuit in Fig. \ref{fig:simplified_qubit_circuit}, transpiled into IBMQ basis gates, requires 1 CNOT gate and 7 single-qubit gates. A recent experimental setting for tomography executed in an ion-trapped quantum processor requires a unitary operator that, when expressed into IBMQ basis gates, requires 2 CNOT gates and 19 single-qubit gates \cite{Stricker2022}. 

Further improvements in the estimation protocol can be addressed. For example, classical shadow tomography strategies can be considered to reduce the number of shots used to estimate a quantum state \cite{Huang2020, Stricker2022}. If the initial state is pure, we can diagonalize the density matrix $\hat{\rho}_0^S$, find the dominant eigenvalue, and take its eigenstate. This corresponds to the pure state $\ket{\hat{\psi}_0^S} = \cos(\theta/2)\ket{0} + e^{i\phi}\sin(\theta/2)\ket{1},$ where the angles $\theta$ and $\phi$ are defined from the vector $\hat{\vec{s}} = (\hat{s}_x, \hat{s}_y, \hat{s}_z)$ represented in spherical coordinates \cite{Nielsen2011}.

\begin{figure}[tb]
    \centering
    \includegraphics[width = 0.99\linewidth]{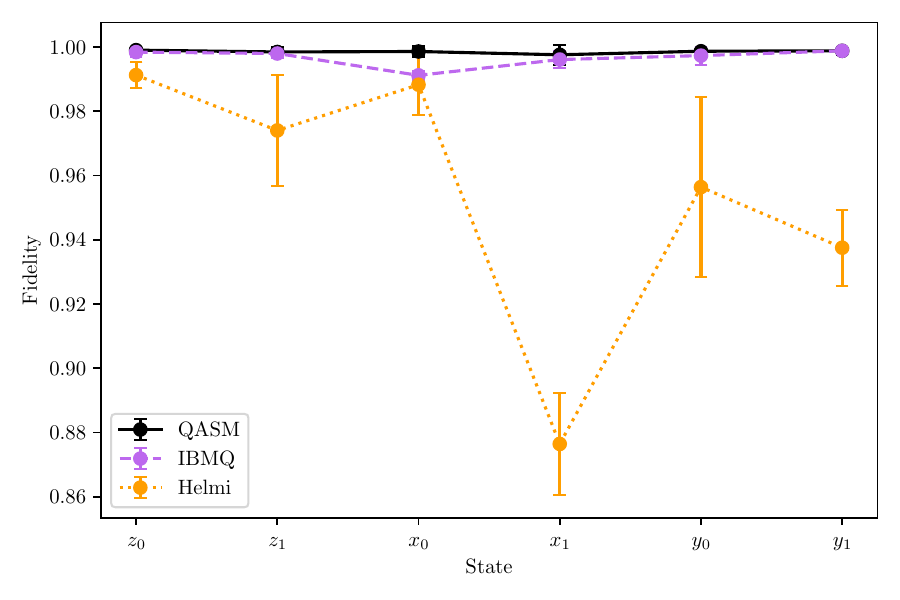}
    \caption{Fidelity of the averaged estimates using the simplified circuit \ref{fig:optimal_circuit_structure} and taking the dominant eigenstate of the density matrix. The estimates were implemented on a QASM simulator (black solid), the IBMQ (purple dashed), and Helmi quantum computers (orange dotted).}
    \label{fig:fidelity_simplified}
\end{figure}
Fig. \ref{fig:fidelity_simplified} shows the fidelity of the estimates using the simplified circuit in Fig. \ref{fig:simplified_qubit_circuit} taking the dominant eigenstate of the estimation. We can see that the estimates have fidelities above 0.97 for IBMQ and above 0.88 on Helmi quantum computer, showing an outstanding performance of the estimation method even in noisy quantum computers.

\section{\label{sec:Conclusions}Conclusions}

Quantum computers have limitations running complex quantum algorithms due to their hardware imperfections.
Assessing the effectiveness of quantum algorithms requires the task of determining the quantum state of a quantum computer via quantum tomography.
Several quantum tomography models have been developed and they can be characterized by the set of POVMs employed to compute an estimate. 
A distinguished set of POVMs corresponds to those that are symmetric informationally complete, due to their symmetric properties and applications in different fields related to quantum information processing.

In this paper, we considered the problem of estimating the quantum state of a qubit when direct measurements are not possible.
To perform the task, we employed the parametric circuit in Fig.~\ref{fig:optimal_circuit_structure} that needs only two CNOT gates to compute complete state estimation of a qubit $S$ by measuring the ancillary qubits $A$ and $B$. 
The optimal parameters $\Theta^*$ that characterize the circuit are determined by optimization of the qTTF~\cite{Rehacek2015}, which is the average of the Fisher error parameter over all possible initial states.
Fig.~\ref{fig:Error_every_state} shows the Fisher error for different sets of parameters, including the optimal set $\Theta^*$. It can be seen that the optimal parameters correspond to estimates with the lowest error and no variations depending on the initial state.
A theoretical analysis of the estimator showed that the optimal parameters $\Theta^*$ correspond to a set of SIC-POVM elements to estimate the qubit $S$. 

Two estimation methods were implemented, the LI estimator and the $R\rho R$ algorithm. According to Fig.~\ref{fig:purity}, LI estimates are occasionally non-physical states while the $R\rho R$ algorithm always generate physical states. However, the latter estimation method is computationally more expensive. Estimates with low purity are the same for LI and $R\rho R$ estimates. To avoid the non-physical estimate problem, we decided to use the $R\rho R$ algorithm.

The demonstrations were implemented on the classical QASM simulator, the \texttt{ibmq\_lima} quantum computer, and the Helmi quantum computer. The results of the Bloch vector components estimates are shown in Table~\ref{tab:qpu_total_s_results_opt}. Demonstrations on QASM simulator show the highest accuracy and precision in the estimates since there is no noise affecting the implementations. Demonstrations on quantum computers showed noisy estimates. Results from the demonstration on IBMQ show higher fidelity than results from Helmi. The demonstrations were also evaluated by considering the fidelity of the estimate.

The proposed circuit allows us to compute estimates that can reach high-fidelity values on real quantum computers. Fig.~\ref{fig:fidelity} shows that demonstrations on IBMQ have fidelities between $0.90$ and $0.99$ despite the hardware errors associated with NISQ devices.
Demonstrations on Helmi show lower accuracy with respect to IBMQ, due to the higher error rates and shorter coherence times of Helmi QPU.

The analytical analysis of the measurement model performed in Sec.~\ref{sec:analytical_analysis} allowed us to simplify the circuit. The POVM elements associated to the circuit with optimal parameters $\Theta^*$ are symmetric informationally complete according to \cite{Renes2004}. Moreover, these SIC-POVM elements can be written in a way such that the gates $U_{A_2},$ $U_{B_1},$ and $U_{B_2}$ are the identity operators while the $U_{A_1}$ gate have angle parameters $\left(-\arccos (1/\sqrt{3}), -\pi/4, 0\right).$ The previous setup was built considering a limitation of measuring the qubit $S$, however, if measurements over $S$ qubit are allowed, the circuit can be reduced to the circuit in Fig.~\ref{fig:simplified_qubit_circuit}. 
When pure states are estimated, it is possible to consider the dominant eigenstate of the estimated density matrix, leading to results with fidelities above 0.97 for IBMQ and 0.88 for Helmi, delivering satisfactory results even on noisy quantum computers. 

It is important to take into account the topology of the QPU, the distribution of the qubits could affect the number of gates used. Here we assumed that the qubit $S$ has two additional qubits besides it as specified in Sec.~\ref{sec:Quantum_hardware_specifications}. Also, There could be cases where the two-qubit basis gate of the QPU is not a CNOT. In that case, the circuit in Fig.~\ref{fig:optimal_circuit_structure} can be modified to include such changes, performing the optimization following the same methodology.
Also, after the measurement, it has to be taken into account that the qubit $S$ does not remain in the same state.
The estimates can be improved by the implementation of error mitigation algorithms that take into account, for example, state preparation and measurement errors, as well as quantum gate errors at different stages of the estimation~\cite{Jattana2020, Johnstun2021}.
\section{Acknowledgements}
We acknowledge the use of IBM Quantum services for this work. The views expressed are those of the authors and do not reflect the official policy or position of IBM or the IBM Quantum team.
Also, we thank the Technical Research Center of Finland VTT for their interest in collaborating as well as the use of the Helmi quantum processing unit. 
We, C. A. Galvis-Florez and Simo Särkkä want to gratefully acknowledge funding from the Academy of Finland project 350221.
Finally, we acknowledge the reviewers of the paper that positively contributed to an improvement in our final results.
\bibliographystyle{IEEEtran}
\bibliography{bibliography.bib}

\begin{thebibliography}{10}
\providecommand{\url}[1]{#1}
\csname url@samestyle\endcsname
\providecommand{\newblock}{\relax}
\providecommand{\bibinfo}[2]{#2}
\providecommand{\BIBentrySTDinterwordspacing}{\spaceskip=0pt\relax}
\providecommand{\BIBentryALTinterwordstretchfactor}{4}
\providecommand{\BIBentryALTinterwordspacing}{\spaceskip=\fontdimen2\font plus
\BIBentryALTinterwordstretchfactor\fontdimen3\font minus
  \fontdimen4\font\relax}
\providecommand{\BIBforeignlanguage}[2]{{%
\expandafter\ifx\csname l@#1\endcsname\relax
\typeout{** WARNING: IEEEtran.bst: No hyphenation pattern has been}%
\typeout{** loaded for the language `#1'. Using the pattern for}%
\typeout{** the default language instead.}%
\else
\language=\csname l@#1\endcsname
\fi
#2}}
\providecommand{\BIBdecl}{\relax}
\BIBdecl

\bibitem{Pelucchi2022}
E.~Pelucchi, G.~Fagas \emph{et~al.}, ``The potential and global outlook of
  integrated photonics for quantum technologies,'' \emph{Nature Reviews
  Physics}, vol.~4, no.~3, pp. 194--208, 2022.

\bibitem{MacFarlane2003}
A.~G.~J. MacFarlane, J.~P. Dowling, and G.~J. Milburn, ``Quantum technology:
  the second quantum revolution,'' \emph{Philosophical Transactions of the
  Royal Society of London. Series A: Mathematical, Physical and Engineering
  Sciences}, vol. 361, no. 1809, pp. 1655--1674, 2003.

\bibitem{Nielsen2011}
M.~A. Nielsen and I.~L. Chuang, \emph{Quantum Computation and Quantum
  Information: 10th Anniversary Edition}, 10th~ed.\hskip 1em plus 0.5em minus
  0.4em\relax USA: Cambridge University Press, 2011.

\bibitem{Wu2021}
Y.~Wu, W.-S. Bao \emph{et~al.}, ``Strong quantum computational advantage using
  a superconducting quantum processor,'' \emph{Phys. Rev. Lett.}, vol. 127, p.
  180501, 2021.

\bibitem{Madsen2022}
L.~S. Madsen, F.~Laudenbach \emph{et~al.}, ``Quantum computational advantage
  with a programmable photonic processor,'' \emph{Nature}, vol. 606, no. 7912,
  pp. 75--81, 2022.

\bibitem{Johnstun2021}
S.~Johnstun and J.-F. Van~Huele, ``Understanding and compensating for noise on
  {IBM} quantum computers,'' \emph{American Journal of Physics}, vol.~89,
  no.~10, pp. 935--942, 2021.

\bibitem{Preskill2018}
J.~Preskill, ``Quantum {C}omputing in the {NISQ} era and beyond,''
  \emph{{Quantum}}, vol.~2, p.~79, 2018.

\bibitem{Bharti2022}
K.~Bharti, A.~Cervera-Lierta \emph{et~al.}, ``Noisy intermediate-scale quantum
  algorithms,'' \emph{Rev. Mod. Phys.}, vol.~94, p. 015004, 2022.

\bibitem{Changjun2020}
C.~Kim, K.~D. Park, and J.-K. Rhee, ``Quantum error mitigation with artificial
  neural network,'' \emph{IEEE Access}, vol.~8, pp. 188\,853--188\,860, 2020.

\bibitem{Jattana2020}
M.~S. Jattana, F.~Jin \emph{et~al.}, ``General error mitigation for quantum
  circuits,'' \emph{Quantum Information Processing}, vol.~19, no.~11, p. 414,
  2020.

\bibitem{Bendersky2009}
A.~Bendersky, F.~Pastawski, and J.~P. Paz, ``Selective and efficient quantum
  process tomography,'' \emph{Phys. Rev. A}, vol.~80, p. 032116, 2009.

\bibitem{James2001}
D.~F.~V. James, P.~G. Kwiat \emph{et~al.}, ``Measurement of qubits,''
  \emph{Phys. Rev. A}, vol.~64, p. 052312, 2001.

\bibitem{Dong2022}
D.~Dong and I.~R. Petersen, ``Quantum estimation, control and learning:
  Opportunities and challenges,'' \emph{Annual Reviews in Control}, vol.~54,
  pp. 243--251, 2022.

\bibitem{Chantasri2021}
A.~Chantasri, I.~Guevara \emph{et~al.}, ``Unifying theory of quantum state
  estimation using past and future information,'' \emph{Physics Reports}, vol.
  930, pp. 1--40, 2021.

\bibitem{Dariano2002}
G.~M. D'Ariano, ``Universal quantum observables,'' \emph{Physics Letters A},
  vol. 300, no.~1, pp. 1--6, 2002.

\bibitem{Mehmani2011}
B.~Mehmani and T.~M. Nieuwenhuizen, ``An overview on single apparatus quantum
  measurements,'' \emph{Journal of Computational and Theoretical Nanoscience},
  vol.~8, no.~6, pp. 937--948, 2011.

\bibitem{Wang2014}
H.~Wang, W.~Zheng \emph{et~al.}, ``Determining an $n$-qubit state by a single
  apparatus through a pairwise interaction,'' \emph{Phys. Rev. A}, vol.~89, p.
  032103, 2014.

\bibitem{Lohani2021}
S.~Lohani, T.~A. Searles \emph{et~al.}, ``On the experimental feasibility of
  quantum state reconstruction via machine learning,'' \emph{IEEE Transactions
  on Quantum Engineering}, vol.~2, pp. 1--10, 2021.

\bibitem{Dominik2022}
D.~Koutn\'y, L.~Motka \emph{et~al.}, ``Neural-network quantum state
  tomography,'' \emph{Phys. Rev. A}, vol. 106, p. 012409, 2022.

\bibitem{Huang2020}
H.-Y. Huang, R.~Kueng, and J.~Preskill, ``Predicting many properties of a
  quantum system from very few measurements,'' \emph{Nature Physics}, vol.~16,
  no.~10, pp. 1050--1057, 2020.

\bibitem{Stricker2022}
R.~Stricker, M.~Meth \emph{et~al.}, ``Experimental single-setting quantum state
  tomography,'' \emph{PRX Quantum}, vol.~3, p. 040310, Oct 2022.

\bibitem{Heinosaari2012}
T.~Heinosaari and M.~Ziman, \emph{The Mathematical Language of Quantum Theory:
  From Uncertainty to Entanglement}.\hskip 1em plus 0.5em minus 0.4em\relax
  Cambridge University Press, 2012.

\bibitem{Flammia2005}
S.~T. Flammia, A.~Silberfarb, and C.~M. Caves, ``Minimal informationally
  complete measurements for pure states,'' \emph{Foundations of Physics},
  vol.~35, no.~12, pp. 1985--2006, 2005.

\bibitem{Rehacek2015}
J.~Řeháček, Y.~S. Teo, and Z.~Hradil, ``Determining which quantum
  measurement performs better for state estimation,'' \emph{Phys. Rev. A},
  vol.~92, p. 012108, 2015.

\bibitem{Renes2004}
J.~M. Renes, R.~Blume-Kohout \emph{et~al.}, ``Symmetric informationally
  complete quantum measurements,'' \emph{Journal of Mathematical Physics},
  vol.~45, no.~6, pp. 2171--2180, 2004.

\bibitem{Tavakoli2020}
A.~Tavakoli, I.~Bengtsson \emph{et~al.}, ``Compounds of symmetric
  informationally complete measurements and their application in quantum key
  distribution,'' \emph{Phys. Rev. Res.}, vol.~2, p. 043122, 2020.

\bibitem{Shang2018}
J.~Shang, A.~Asadian \emph{et~al.}, ``Enhanced entanglement criterion via
  symmetric informationally complete measurements,'' \emph{Phys. Rev. A},
  vol.~98, p. 022309, 2018.

\bibitem{Hacohen-Gourgy2016}
S.~Hacohen-Gourgy, L.~S. Martin \emph{et~al.}, ``Quantum dynamics of
  simultaneously measured non-commuting observables,'' \emph{Nature}, vol. 538,
  no. 7626, pp. 491--494, Oct 2016.

\bibitem{Saavedra2019}
D.~Saavedra and K.~M. Fonseca-Romero, ``Complete and incomplete state
  estimation via the simultaneous unsharp measurement of two incompatible qubit
  operators,'' \emph{Phys. Rev. A}, vol.~99, p. 042130, 2019.

\bibitem{Galvis2023}
C.~A. Galvis~Florez, J.~Martínez-Cifuentes, and K.~M. Fonseca-Romero,
  ``Partial and complete qubit estimation using a single observable:
  optimization and quantum simulation,'' 2023,
  \href{https://arxiv.org/abs/2301.11121}{https://arxiv.org/abs/2301.11121},Accessed:
  22-04-2023.

\bibitem{Peres1986}
A.~Peres, ``When is a quantum measurement?'' \emph{American Journal of
  Physics}, vol.~54, no.~8, pp. 688--692, 1986.

\bibitem{Perarnau2017}
M.~Perarnau-Llobet and T.~M. Nieuwenhuizen, ``Simultaneous measurement of two
  noncommuting quantum variables: Solution of a dynamical model,'' \emph{Phys.
  Rev. A}, vol.~95, p. 052129, 2017.

\bibitem{Paris2004}
M.~Paris and J.~Řeháček, \emph{Quantum State Estimation}.\hskip 1em plus
  0.5em minus 0.4em\relax Springer, Berlin, Heidelberg, 2004.

\bibitem{Teo2015}
Y.~S. Teo, \emph{Introduction to Quantum-State Estimation}.\hskip 1em plus
  0.5em minus 0.4em\relax World Scientific, 2015.

\bibitem{Rehacek2007}
J.~Řeháček, Z.~Hradil \emph{et~al.}, ``Diluted maximum-likelihood algorithm
  for quantum tomography,'' \emph{Phys. Rev. A}, vol.~75, p. 042108, 2007.

\bibitem{Fiurasek2001}
J.~Fiurás\v{e}k, ``Maximum-likelihood estimation of quantum measurement,''
  \emph{Phys. Rev. A}, vol.~64, p. 024102, 2001.

\bibitem{Kay1993}
S.~M. Kay, \emph{Fundamentals of Statistical Signal Processing: Estimation
  Theory}.\hskip 1em plus 0.5em minus 0.4em\relax Prentice Hall, 1993.

\bibitem{2020SciPy-NMeth}
P.~Virtanen, R.~Gommers \emph{et~al.}, ``{{SciPy} 1.0: Fundamental Algorithms
  for Scientific Computing in Python},'' \emph{Nature Methods}, vol.~17, pp.
  261--272, 2020.

\bibitem{Rehacek2004}
J.~Řeháček, B.-G. Englert, and D.~Kaszlikowski, ``Minimal qubit
  tomography,'' \emph{Phys. Rev. A}, vol.~70, p. 052321, Nov 2004.

\bibitem{Qiskit}
M.~Treinish, J.~Gambetta \emph{et~al.} (2023) Qiskit.
  \href{https://doi.org/10.5281/zenodo.7757946}{https://doi.org/10.5281/zenodo.7757946}.
  Accessed: 27-04-2023.

\bibitem{Devitt2016}
S.~J. Devitt, ``Performing quantum co<mputing experiments in the cloud,''
  \emph{Phys. Rev. A}, vol.~94, p. 032329, 2016.

\bibitem{ibm-quantum}
{IBM Quantum}. (2021)
  \href{https://quantum-computing.ibm.com/}{https://quantum-computing.ibm.com/}.
  Accessed: 27-04-2023.

\bibitem{FiQCI}
{Finnish Quantum Computing Infrastructure}. (2022)
  \href{https://fiqci.fi/}{https://fiqci.fi/}. Accessed: 27-04-2023.

\bibitem{Haah2017}
J.~Haah, A.~W. Harrow \emph{et~al.}, ``Sample-optimal tomography of quantum
  states,'' \emph{IEEE Transactions on Information Theory}, vol.~63, no.~9, pp.
  5628--5641, 2017.

\bibitem{Garcia-Perez2021}
G.~Garc\'{\i}a-P\'erez, M.~A. Rossi \emph{et~al.}, ``Learning to measure:
  Adaptive informationally complete generalized measurements for quantum
  algorithms,'' \emph{PRX Quantum}, vol.~2, p. 040342, Nov 2021.

\bibitem{Jiang2020}
Z.~Jiang, A.~Kalev \emph{et~al.}, ``Optimal fermion-to-qubit mapping via
  ternary trees with applications to reduced quantum states learning,''
  \emph{{Quantum}}, vol.~4, p. 276, Jun. 2020.

\end{thebibliography}
\end{document}